\documentclass[preprint,aps]{revtex4}
\usepackage{graphicx}
\begin{document}
\title{Determination of pair correlation function in a weakly correlated weakly inhomogeneous plasma}
\author{Anirban Bose}
\affiliation{ Serampore College, Serampore, Hooghly, India}

\begin{abstract}
In a weakly correlated inhomogeneous plasma an equation of pair correlation function  is obtained utilizing the Bogoliubov-Born-Green-Kirkwood-Yvon (BBGKY)  hierarchy of equations. In this article the pair correlation function has been calculated from the same equation in the weakly inhomogeneous limit by using a perturbative theory.
\end{abstract}

\maketitle

The importance of pair correlation function in plasma systems has been recognised for a long time. It plays a crucial role in the study of thermodynamic properties of the concerned systems and is of fundamental importance to determine the collision integral of the Fokker-Planck equation.  Consequently, determination and study of pair correlation function has been the topic of discussion since long and even at present it is able to attract the attention of the modern day researchers.
Dupree \cite{kn:dup} has developed an elegant method for determining the pair correlation function from the one-particle distribution function when it varies slowly in space and time. The technique due to Dupree has been used to determine the pair correlation function in a dilute, uniform, quiescent plasma \cite{kn:wolf}. O'Neil and Rostoker have calculated the three-body and two-body electron correlation functions for a homogeneous plasma in thermal equilibrium \cite{kn:neil}. Jancovici \cite{kn:jan} has investigated the short-range behavior of the pair correlation function in a dense one component plasma (jellium) and calculated the quantum pair correlation function  by treating the many-body quantum effects by a perturbation theory, and by using a semiclassical approximation based on path integrals. The pair-correlation function for a one-component, two-dimensional classical plasma is investigated by  C Deutsch and M Lavaudcite \cite{kn:cd} within the framework of the Debye approximation along the lines of the Cohen-Murphy method. The canonical equilibrium properties of classical Coulomb systems are investigated for any value of the plasma parameter \cite{kn:fur}, through the nodal expansion of the two-particle correlation function. The pair distribution function is obtained by the Debye-H$\ddot{u}$ckel theory for the weakly correlated system. On the other side, the triplet and higher order correlation should be included in case of strongly coupled systems which can be determined from the integral equation technique discussed in equilibrium statistical mechanics \cite{kn:han}. Ornstein Zernike equation is one of those which has been successfully applied in several cases with some closure relations. The integral equations approach has been used to calculate charge, screening radius and the interaction potential of dust particles. This approach is based on experimentally obtained pair correlation functions \cite{kn:fortov}. Molecular dynamics simulations has been performed for the pair correlation functions of strongly coupled plasmas with species of unequal temperature \cite{kn:nrs}. Recently, parametrization of the pair correlation function and the static structure factor of the Coulomb one component plasma (OCP) has been done from the weakly coupled regime to the strongly coupled regime \cite{kn:des}.
In a previous paper \cite{kn:ab} an equation of pair correlation has been obtained from the first two members of BBGKY hierarchy involving the one-particle and two-particle distribution function in a weakly correlated inhomogeneous plasma system under certain approximations and the next step is to derive the pair correlation function from that equation under different physical conditions. In this Brief Communication pair correlation function is derived from the same equation for a weakly correlated inhomogeneous plasma system.

We have considered a plasma of N electrons and N infinitely massive ions which are randomly distributed in a volume V.
The plasma is in thermal equilibrium and the electron-electron pair correlation function $g_{12}$ is written according to the following expression
\begin{eqnarray}
 g_{12}({\bf {X_1,X_2}})&=&f_{1}({\bf X_1})f_{1}({\bf X_2})\chi_{12}({\bf x_1,x_2})
\label{v33}\end{eqnarray}

The single particle (electrons) distribution functions $f_{1}({\bf X_1})$
and $ f_{1}({\bf X_2})$ are functions of both position and
velocity. $\chi_{12}$ is an unknown symmetric function of the position coordinates $\bf{x}_{1}$ and $\bf{x}_{2}$.

As derived earlier \cite{kn:ab} for a weakly correlated dust particles, the equation of $\chi_{12}$ (for the dust particles in the limit $\chi_{12}\ll1$) is \cite{kn:ab}

\begin{eqnarray}
\chi_{12}=-\frac{\phi_{12}}{k_{B}T}- \frac{1}{k_{B}T}\int d{\mathbf{x_{3}}}
\phi_{13}n(3)\chi_{23}
\label{v1}\end{eqnarray}
where $\phi_{ij}$ is the energy of Coulomb interaction between the ith and jth dust particle.

In this article, the same equation is applied to the weakly correlated electrons. In addition to that, the system is considered to be weakly inhomogeneous. The inhomogeneous density of electrons is considered as
\begin{eqnarray}
n(3)= n_{0}+n_{1}(3) 
\end{eqnarray}
where $n_{0}$ is the homogeneous part and $n_{1}(3)$ is the inhomogeneous perturbation.

Therefore, eq.(\ref{v1}) is expressed as
\begin{equation}
\chi_{12}^{0}+\chi_{12}^{1}=-\frac{\phi_{12}}{k_{B}T}- \frac{1}{k_{B}T}\int d{\mathbf{x_{3}}}
\phi_{13}(n_{0}+n_{1}(3))(\chi_{23}^{0}+\chi_{23}^{1})
\end{equation}
where $\chi_{ij}^{0}$ and $\chi_{ij}^{1}$ are the zeroth and first order solution of the pair correlation respectively.

The zeroth order equation is
\begin{eqnarray}
\chi_{12}^{0}=-\frac{\phi_{12}}{k_{B}T}- \frac{1}{k_{B}T}\int d{\mathbf{x_{3}}}
\phi_{13}n_{0}\chi_{23}^{0}
\end{eqnarray}
Hence,
\begin{eqnarray}
\chi_{12}^{0}=-A\frac{e^{-k_{D}r}}{r}
\end{eqnarray}
where, $r=\mid \textbf{x}_{1}-\textbf{x}_{2}\mid$ and A=$\frac{e^{2}}{k_{B}T￢ﾀﾢ}$. The electronic charge is e and T is the absolute temperature of the electrons.

The first order equation is  
\begin{eqnarray}
\chi_{12}^{1}+ \frac{1}{k_{B}T}\int d{\mathbf{x_{3}}}
\phi_{13}n_{0}\chi_{23}^{1}=- \frac{1}{k_{B}T}\int d{\mathbf{x_{3}}}
\phi_{13}n_{1}(3)\chi_{23}^{0}
\label{v6}\end{eqnarray}
The general expression of $\chi^{1}_{ij}$ may be chosen as
\begin{eqnarray}
\chi_{ij}^{1}=\theta(\textbf{x}_{i}-\textbf{x}_{j})g(\frac{\textbf{x}_{i}+\textbf{x}_{j}￢ﾀﾢ}{2￢ﾀﾢ})
\end{eqnarray}
where $\theta(\textbf{x}_{i}-\textbf{x}_{j})$ is a function of $\mid\textbf{x}_{i}-\textbf{x}_{j}\mid$.

Therefore, eq.(\ref{v6}) is expressed as

\begin{eqnarray} 
\theta(\textbf{x}_{1}-\textbf{x}_{2})g(\frac{\textbf{x}_{1}+\textbf{x}_{2}￢ﾀﾢ}{2￢ﾀﾢ})+ \frac{1}{k_{B}T}\int d{\mathbf{x_{3}}}
\phi_{13}n_{0}\theta(\textbf{x}_{2}-\textbf{x}_{3})g(\frac{\textbf{x}_{2}+\textbf{x}_{3}￢ﾀﾢ}{2￢ﾀﾢ})=
\nonumber\end{eqnarray}
\begin{eqnarray}
 - \frac{1}{k_{B}T}\int d{\mathbf{x_{3}}}
\phi_{13}n_{1}(3)\chi_{23}^{0}
\label{v8}\end{eqnarray}

Using the following transformations,
$$g(\frac{\mathbf{x}_{i}+\mathbf{x}_{j}￢ﾀﾢ}{2￢ﾀﾢ})=\int g(\mathbf{q})e^{i\mathbf{q}\cdot(\frac{\mathbf{x}_{i}+\mathbf{x}_{j}￢ﾀﾢ}{2￢ﾀﾢ})}d\mathbf{q}$$
$$
 \frac{1}{\mid
 {\bf{x_i-x_j}}\mid}
=\int d{\mathbf{k}}
\frac{1}{k^{2}}e^{i\mathbf{k}\cdot(\mathbf{x}_{i}-\mathbf{x}_{j})}\nonumber
$$
$$\theta(\textbf{x}_{i}-\textbf{x}_{j})=\int \theta(\textbf{S})e^{i\mathbf{S}\cdot(\mathbf{x}_{j}-\mathbf{x}_{i})}d\textbf{S}$$

the left hand side of eq.(\ref{v8}) is given by 
\begin{eqnarray}
\theta(\textbf{x}_{1}-\textbf{x}_{2})g(\frac{\textbf{x}_{1}+\textbf{x}_{2}￢ﾀﾢ}{2￢ﾀﾢ})+ \frac{n_{0}}{k_{B}T}\int d{\mathbf{x_{3}}}
\phi_{13}\theta(\textbf{x}_{2}-\textbf{x}_{3})g(\frac{\textbf{x}_{2}+\textbf{x}_{3}￢ﾀﾢ}{2￢ﾀﾢ})=
\nonumber\end{eqnarray}
\begin{eqnarray}
\int\frac{\mid\mathbf{S}+\frac{\mathbf{q}}{2}\mid^{2}+{k}_{D}^{2}}{\mid\mathbf{S}+\frac{\mathbf{q}}{2}\mid^{2}}\theta(\textbf{S})e^{i\mathbf{S}\cdot(\mathbf{x}_{1}-\mathbf{x}_{2})}g(\textbf{q})e^{i\textbf{q}\cdot(\frac{\mathbf{x}_{1}+\mathbf{x}_{2}￢ﾀﾢ}{2￢ﾀﾢ})}d\textbf{S}d\textbf{q}
\label{v9}\end{eqnarray}

Let, $n_{1}(3)=B\cos (\textbf{p}\cdot \textbf{x}_{3})$.  Hence, the right hand side of eq.(\ref{v8}) is
\begin{equation}
-\frac{1}{k_{B}T￢ﾀﾢ}\int\phi_{13}n_{1}(3)\chi_{23}^{0}d\mathbf{x}_{3}=\frac{ABe^{2}}{k_{B}T}\int\frac{e^{i\mathbf{k}_{1}\cdot(\mathbf{x}_{1}-\mathbf{x}_{3})}}{\mathbf{k}_{1}^{2}}
\cos(\mathbf{p}\cdot
\mathbf{x}_{3})\frac{e^{i\mathbf{k}_{3}\cdot(\mathbf{x}_{2}-\mathbf{x}_{3})}}{\mathbf{k}_{3}^{2}+{k}_{D}^{2}}d\mathbf{k}_{1}d\mathbf{k}_{3}d\mathbf{x}_{3}\end{equation}

\begin{equation}
=\frac{ABe^{2}}{2k_{B}T}\int\frac{e^{i\mathbf{k}_{1}\cdot\mathbf{x}_{1}}}{\mathbf{k}_{1}^{2}}
[\delta(\mathbf{p}-\mathbf{k}_{1}-\mathbf{k}_{3})+\delta(\mathbf{p}+\mathbf{k}_{1}+\mathbf{k}_{3})]\frac{e^{i\mathbf{k}_{3}\cdot\mathbf{x}_{2}}}{\mathbf{k}_{3}^{2}+{k}_{D}^{2}}d\mathbf{k}_{1}d\mathbf{k}_{3}\end{equation}
\begin{equation}
=\frac{ABe^{2}}{2k_{B}T}\int\frac{e^{i\mathbf{k}_{1}\cdot\mathbf{x}_{1}}}{\mathbf{k}_{1}^{2}}\frac{e^{i(\mathbf{p}-\mathbf{k}_{1})\cdot\mathbf{x}_{2}}}{\mid\mathbf{k}_{1}-\mathbf{p}\mid^{2}+{k}_{D}^{2}}
d\mathbf{k}_{1}+\frac{ABe^{2}}{2k_{B}T}\int\frac{e^{i\mathbf{k}_{4}\cdot\mathbf{x}_{1}}}{\mathbf{k}_{4}^{2}}\frac{e^{-i(\mathbf{p}+\mathbf{k}_{4})\cdot\mathbf{x}_{2}}}{\mid\mathbf{k}_{4}+\mathbf{p}\mid^{2}+{k}_{D}^{2}}
d\mathbf{k}_{4}\end{equation}
Substituting,
$$\mathbf{k}_{1}-\mathbf{p}/2=\mathbf{q}$$
$$\mathbf{k}_{4}+\mathbf{p}/2=\mathbf{u}$$
\begin{equation}
=\frac{ABe^{2}}{2k_{B}T}\int\frac{e^{i\mathbf{q}\cdot(\mathbf{x}_{1}-\mathbf{x}_{2})}}{\mid
\mathbf{q}+\frac{\mathbf{p}}{2}\mid^{2}}\frac{e^{i\frac{\mathbf{p}}{2}\cdot(\mathbf{x}_{1}+\mathbf{x}_{2})}}{\mid\mathbf{q}-\frac{\mathbf{p}}{2}\mid^{2}+{k}_{D}^{2}}
d\mathbf{q}+\frac{ABe^{2}}{2k_{B}T}\int\frac{e^{i\mathbf{u}\cdot(\mathbf{x}_{1}-\mathbf{x}_{2})}}{\mid
\mathbf{u}-\frac{\mathbf{p}}{2}\mid^{2}}\frac{e^{-i\frac{\mathbf{p}}{2}\cdot(\mathbf{x}_{1}+\mathbf{x}_{2})}}{\mid\mathbf{u}+\frac{\mathbf{p}}{2}\mid^{2}+{k}_{D}^{2}}
d\mathbf{u}\label{v14}\end{equation}

Comparing eq.(\ref{v9}) and eq.(\ref{v14}), we may identify
\begin{equation}
g(\textbf{q})=\frac{\delta(\textbf{q}-\textbf{p})+\delta(\textbf{q}+\textbf{p})￢ﾀﾢ}{2￢ﾀﾢ}
\end{equation}
Hence,
\begin{equation} 
g(\frac{\textbf{x}_{1}+\textbf{x}_{2}￢ﾀﾢ}{2￢ﾀﾢ})=B\cos[\frac{\mathbf{p}}{2}\cdot(\mathbf{x}_{1}+\mathbf{x}_{2})]
\end{equation}
and
\begin{equation}
\theta(\textbf{S})=\frac{Ak_{D}^{2}}{n_{0}(\mid\mathbf{S}+\frac{\mathbf{p}}{2}\mid^{2}+{k}_{D}^{2})(\mid\mathbf{S}-\frac{\mathbf{p}}{2}\mid^{2}+{k}_{D}^{2})}
\label{v17}\end{equation}
Hence,
\begin{equation}
\theta(\textbf{x}_{1}-\textbf{x}_{2})=\int\frac{Ak_{D}^{2}e^{i\mathbf{S}\cdot(\mathbf{x}_{1}-\mathbf{x}_{2})}}{n_{0}(\mid\mathbf{S}+\frac{\mathbf{p}}{2}\mid^{2}+{k}_{D}^{2})(\mid\mathbf{S}-\frac{\mathbf{p}}{2}\mid^{2}+{k}_{D}^{2})}d\textbf{S}
\end{equation}
In the weakly inhomogeneous limit, if terms of the order of $(p/k_{D})^{2}$ and subsequent higher orders are neglected
\begin{equation}
\theta(\textbf{S})=\frac{Ak_{D}^{2}}{n_{0}(\mid\mathbf{S}\mid^{2}+{k}_{D}^{2})^{2}}
\end{equation}
\begin{equation}
\chi_{12}^{1}=AB\frac{k_{D}}{2n_{0}}e^{-k_{D}r}\cos[\frac{\mathbf{p}}{2}\cdot(\mathbf{x}_{1}+\mathbf{x}_{2})]
\end{equation}
Therefore,
\begin{equation}
\chi_{12}=-A\frac{e^{-k_{D}r}}{r}+AB\frac{k_{D}}{2n_{0}}e^{-k_{D}r}\cos[\frac{\mathbf{p}}{2}\cdot(\mathbf{x}_{1}+\mathbf{x}_{2})]
\end{equation}
\begin{equation}
=-A\frac{e^{-k_{D}r}}{r}(1-B\frac{k_{D}r}{2n_{0}}\cos[\frac{\mathbf{p}}{2}\cdot(\mathbf{x}_{1}+\mathbf{x}_{2})])
\label{v23}\end{equation}

From the above expression one may conclude that in the presence of inhomogeneity in the form of sinusoidal structure as a perturbation to the homogeneous system the pair correlation function gets modified from the Debye-H$\ddot{u}$ckel form which is previously derived in the case of homogeneous system. If the homogeneity of the system is restored (B=0), eq.(\ref{v23}) correctly gives back the Debye-H$\ddot{u}$ckel form. On the contrary, the inclusion of the inhomogeneity modifies the expression and the correlation does not only depend on the separation of the concerned partcles but also on the position where the pair is placed. For instance, when the position of the pair is near the maximums of the cosine function the $AB\frac{k_{D}}{2n_{0}}e^{-k_{D}r}$ is added to the Debye-Huckel form but if the pair with the same separation is placed near the positions where the value of the cosine function is zero, nothing is added to the Debye-Huckel form and $AB\frac{k_{D}}{2n_{0}}e^{-k_{D}r}$ is subtracted in case of the minimums of the same cosine function. The results clearly indicate that the magnitude of the correlation keeps changing with the positioning of the pair even if the separation between the particles remaining the same.

The above expression also demonstrates the directionality factor in the pair correlation. For a fixed particle even for the same separation the strength of correlation depends on the relative position of the second particle with respect to that fixed particle.  This is the consequence of the argument of the cosine function. The dot product $\frac{\mathbf{p}}{2}\cdot(\mathbf{x}_{1}+\mathbf{x}_{2})$ introduces the directionality factor in the pair correlation. Finally we could include the higher order  terms of the expansion in eq.(\ref{v17}) to enhance the inhomogeneity of the plasma systems within the weakly inhomogeneous limits. Even the complete solution can be obtained by calculating the inverse fourier transformation of $\theta$  as given by eq.(\ref{v17}). At the end it should be mentioned that the above pair correlation is meaningful only in the weakly inhomogeneous limit and $\chi_{ij}\ll 1$.

The research is supported by the Department of Science and Technology, West Bengal (File No. ST/P/S$\&$T/9G-36/2017).

\end{document}